\documentclass[a4paper]{jpconf}
\usepackage{graphicx}
\usepackage{lineno}

\begin{document}

\title{Experimental results on $t\bar{t}+W/Z/\gamma$ and SM top couplings from the Tevatron and the LHC}

\author{Tamara V\'azquez Schr\"oder for the ATLAS, CDF, CMS and D0 Collaborations}

\address{II. Physikalisches Institut, Georg-August Universit\"at G\"ottingen, G\"ottingen, Germany}

\ead{tamara.vazquez-schroeder@phys.uni-goettingen.de}

\begin{abstract}
Experimental results from the CDF and D0 Collaborations at the Tevatron and the ATLAS and CMS Collaborations at the LHC on the processes related to probing top quark couplings are presented. Evidence of both $t\bar{t}Z$ and $t\bar{t}W$ processes is reported. All measurements are in agreement with the SM expectations.
\end{abstract}

\section{Motivation}
The top quark was discovered in 1995 by the CDF and D0 Collaborations \cite{D0:topdisc,CDF:topdisc}. It couples to the Standard Model fields through its gauge and Yukawa interactions. Some of these couplings have been investigated at the Tevatron, through studies of the $Wtb$ vertex and the $t\bar{t}\gamma$ production, while others, such as the $t\bar{t}Z$ and $t\bar{t}H$ production, are becoming accessible only with the high statistics top quark sample at the LHC, also called for this reason a `top quark factory'. At hadron colliders, the first evidence of the coupling of the top quark to the $\gamma, Z,$ and $H$ boson will come from the production rate, while constraints on the coupling of the top quark with the $W$ boson come from both the top quark decay and the single top production. Given its large mass, the top quark may play a special role in the electroweak symmetry breaking (EWSB) and therefore, new physics related to EWSB may be found first in top quark precision measurements. Possible new physics signals would cause deviations of the top quark couplings $tZ$, $t\gamma$, and $Wtb$, from the SM prediction.

\section{Wtb coupling}
Information on the coupling of the top quark to the $W$ boson can be obtained from the top quark decay and electroweak single top production.

\subsection{W-helicity measurements}

Since the top quark decays almost exclusively as $t\rightarrow W^{+}b$, the measurement of the $W$ boson helicity in top decays probes the structure of the $Wtb$ vertex, which in the Standard Model (SM) is V-A. Since $W$ bosons are produced as on-shell particles in top quark decays, their polarisation can be longitudinal, left-handed or right-handed. The fractions with a certain polarisation, $F_{0}, F_{L}$ and $F_{R}$, can be extracted from measurements of the angular distribution of the decay products of the top quark, given by:
\begin{linenomath}
	\begin{equation}
		\frac{1}{\sigma}\frac{d\sigma}{d\cos{\theta^{*}}} = \frac{3}{4}(1-\cos^{2}{\theta^{*}})F_{0} + \frac{3}{8}(1-\cos{\theta^{*}})^{2}F_{L} + \frac{3}{8}(1+\cos{\theta^{*}})^{2}F_{R},
	\end{equation}
\end{linenomath}
	where $\theta^{*}$ is defined as the angle between the $W$ boson momentum in the top quark rest frame and the momentum of the down-type decay fermion in the rest frame of the $W$ boson. The next-to-next-to-leading-order (NNLO) QCD prediction for the helicity fractions in the SM, for a top quark mass $m_{t}=172.8$~GeV and a $b$-quark mass $m_{b}=4.8$~GeV, are $F_{0} = 0.687 \pm 0.005$, $F_{L} = 0.311 \pm 0.005$ and $F_{R}=0.0017 \pm 0.0001$ \cite{TheoryNNLO:Whelicity}. 
	Measurements of the $W$ boson helicity fractions have been performed by both, CDF and D0 experiments \cite{CDF:Whelicity, D0:Whelicity, CDFD0:WhelicityComb}. The CDF measurement is performed with the full dataset of 8.7 \ifb\ in the $\ell$+jets channel using the matrix element method, while the D0 measurement is performed with 5.4 \ifb\ in both, the $\ell$+jets and dilepton channels, using the template method. Both measurements extract simultaneously the $F_{0}$ and $F_{R}$ helicity fractions and determine the $F_{L}$ from unitarity, obtaining values consistent with the SM expectations.
    The ATLAS~\cite{ATLAS:detector} and CMS~\cite{CMS:detector} Collaborations combined their $W$-helicity measurements at $\sqrt{s}=7$ TeV with $35$~\ipb\ and $1.04$~\ifb\ in ATLAS, and $2.2$~\ifb in CMS, measuring simultaneously $F_{0}$, $F_{L}$, and the \ttbar\ normalisation \cite{ATLASCMS:WhelicityComb}. Limits were set on the real part of the $Wtb$ parameters $g_{L}$ and $g_{R}$, which enter in the $Wtb$ effective Lagrangian as:
\begin{linenomath}
	\begin{equation}
		\mathcal{L}=-\frac{g}{\sqrt{2}}\bar{b}\gamma^{\mu}(V_{L}P_{L}+V_{R}P_{R})tW_{\mu}^{-} - \frac{g}{\sqrt{2}}\bar{b}\frac{i\sigma^{\mu\nu}q_{\nu}}{M_{W}}(g_{L}P_{L}+g_{R}P_{R})tW_{\mu}^{-} + \mbox{h.c.}
	\end{equation}
\end{linenomath}
	At tree level, the SM predicts a $V_{L}=1$ and $V_{R}=g_{L}=g_{R}=0$. The measured limits are consistent with the V-A structure of the $Wtb$ vertex. 
	The CMS Collaboration recently performed a measurement at $\sqrt{s}=7$~TeV in the $\ell$+jets channel and a measurement at $\sqrt{s}=8$~TeV in the $\mu$+jets channel with the full dataset of $5.0$ \ifb\ and $19.6$ \ifb, respectively. Both measurements use the reweighting method and measure $F_{0}$, $F_{L}$ and the normalisation of \ttbar\ \cite{ CMS:Whelicity7, CMS:Whelicity8}. In the 7 TeV measurement, limits are also set on the $g_{L}$ and $g_{R}$ parameters, with better sensitivity than the limits from the ATLAS and CMS combined result, and can be seen in Figure~\ref{WhelicityCMS7}. All results are in agreement with the SM predictions.
	
\begin{figure}[h]
\centering
\includegraphics[width=18pc]{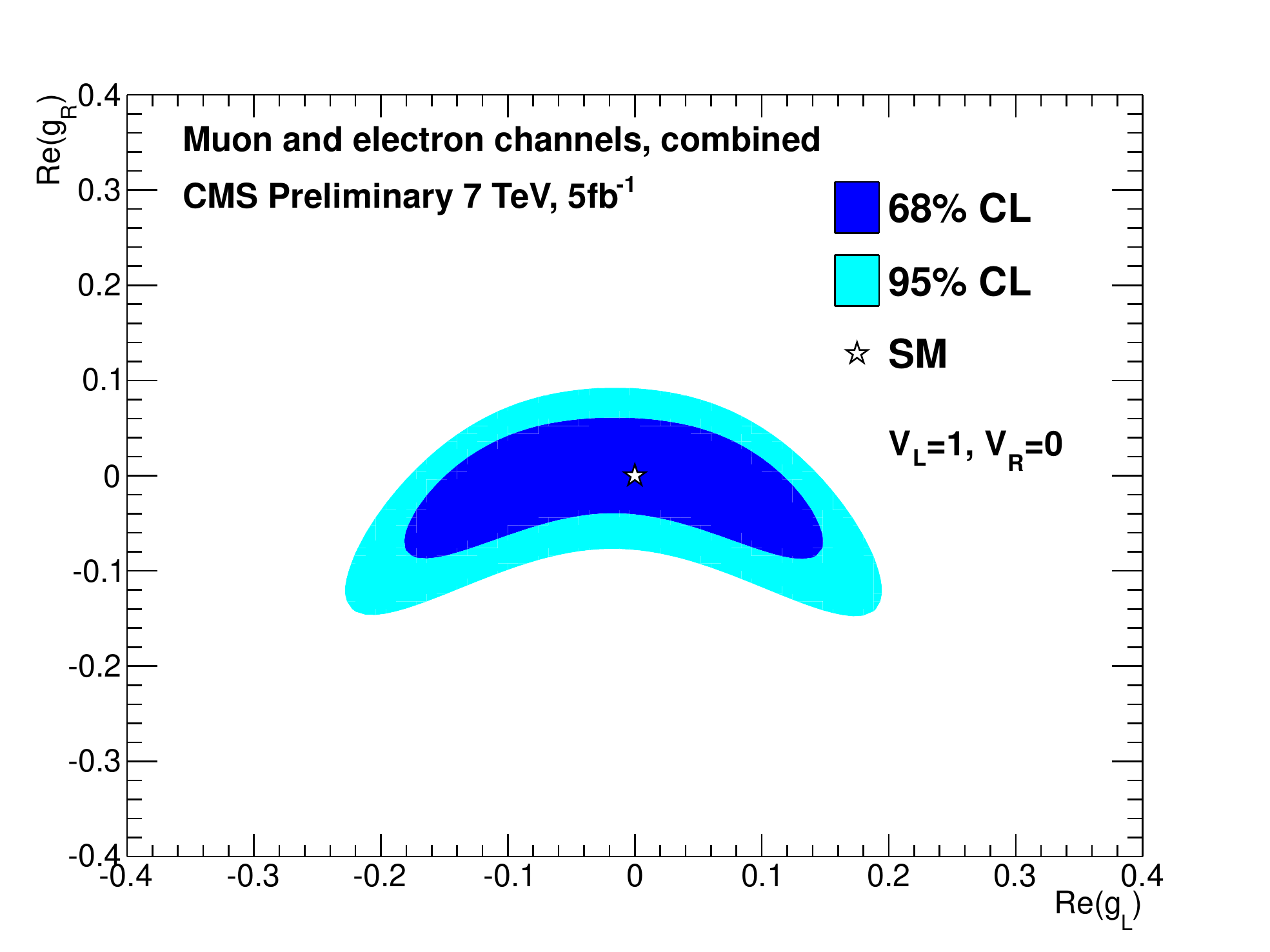}
\caption{\label{WhelicityCMS7} Limits on the real components of the anomalous couplings $g_{L}$ and $g_R$ at 68\% and 95\% CL, for $V_L=1$ and $V_R=0$, performed by the CMS Collaboration at $\sqrt{s}=7$ TeV with the full dataset of $5.0$ \ifb\ \cite{CMS:Whelicity7}.}
\end{figure}	

\subsection{$R_b$ measurements}

Under the assumption of a unitary $3\times 3$ CKM matrix, the top quark decays almost exclusively to $Wb$, i.e. $|V_{tb}| \approx 1$. A fourth generation of quarks would accommodate significantly smaller values of $|V_{tb}|$, affecting, for example, the decay rates in the $t\bar{t}$ production channel. Therefore, a measurement of the ratio of branching fractions of the form:
\begin{linenomath}
	\begin{equation}
		R_{b}= \frac{B(t \rightarrow Wb)}{B(t \rightarrow Wq)} =  \frac{|V_{tb}|^{2}}{|V_{tb}|^{2} + |V_{ts}|^{2} + |V_{td}|^{2}},
 	\end{equation}
\end{linenomath}
would test the three generations assumption. A measurement of $|V_{tb}|$ can also be extracted from $R_{b}$ by assuming a unitary $3\times 3$ CKM matrix, where $R_{b} = |V_{tb}|^{2}$. The most precise measurement to date of $R_{b}$ has been performed by the CMS Collaboration at $\sqrt{s}=8$ TeV in the dilepton channel using a profile likelihood fit with 36 event categories, defined by the lepton flavour, the jet multiplicity, and $b$-tagged jet multiplicity. This measurement results in an unconstrained measured value of $R_{b} = 1.014 \pm 0.003 \mbox{(stat.)} \pm 0.032 \mbox{(syst.)}$, which translates into $|V_{tb}|=1.007 \pm 0.016$ under the three-generation CKM matrix assumption, and a lower limit of $|V_{tb}|> 0.975$ at $95\%$ CL when requiring $|V_{tb}| \le 1$, all consistent with the SM predictions \cite{CMS:Rb}. 

\section{\ttgamma\ measurements} 
At hadron colliders, a measurement of the $t\bar{t}\gamma$ coupling via $q\bar{q} \rightarrow \gamma^{*} \rightarrow t\bar{t}$ is unrealistic due to the overwhelming contribution from QCD processes. Therefore, a more feasible approach to probe the $t\bar{t}\gamma$ coupling is via the measurement of associated production of a photon with a top quark pair. The photon can be radiated via several mechanisms: from the top quark, $pp\rightarrow t\bar{t}\gamma$, or from the top quark decay products, $pp \rightarrow t\bar{t}$ and $t\rightarrow Wb\gamma$. Only events of the first type are sensitive to the $t\bar{t}\gamma$ coupling, and therefore, to the top quark electric charge. However, for a well-defined $t\bar{t}\gamma$ final state, all interferences between both processes have to be taken into account.

First evidence of the associated production of photon radiation with a top quark pair was announced by the CDF Collaboration at $\sqrt{s} = 1.96$ TeV with 6.0 \ifb\ \cite{CDF:ttgamma}. \ttgamma\ events are selected by requiring a standard \ttbar\ $\ell+$jets selection, with one additional central photon with a $p_{T} > 10$ GeV, measuring $\sigma(\ttgamma) = 0.18 \pm 0.07 \mbox{(stat.)} \pm 0.04 \mbox{(syst.)} \pm 0.01 \mbox{(lumi)}$ pb. The ATLAS Collaboration performed a \ttgamma\ measurement at $\sqrt{s} = 7$ TeV with 1.04 \ifb\ in the $\ell+$jets channel, and requiring at least 1 good photon with a $p_{T} > 15$ GeV. The template fit method is used with a photon track isolation distribution, defined as the scalar sum of the transverse momentum of all tracks with $\Delta R < 0.2$ around the photon candidate, as the discriminating variable. The measured \ttgamma\ production cross section for a photon with $p_{T} > 8$~GeV is $\sigma(\ttgamma) = 2.0 \pm 0.5 \mbox{(stat.)} \pm 0.7 \mbox{(syst.)} \pm 0.08 \mbox{(lumi)}$ pb, with an observed significance of 2.7 $\sigma$  \cite{ATLAS:ttgamma}. In both CDF and ATLAS measurements, the main background contribution originates from hadron fakes, and the dominant systematic uncertainty is due to the photon identification efficiency. 
 With a similar strategy to ATLAS, the CMS Collaboration measured the \ttgamma\ production cross section at $\sqrt{s} = 8$ TeV with the full dataset of 19.7 \ifb\ in the $\mu+$ jets channel. \ttbar\ events with at least one good central photon with $E_{T}> 25$ GeV and specific isolation requirements are selected. The discriminating variable used in this case is the charged hadron isolation of the photon candidate, motivated by the dominant background contribution of hadron fakes. The measured \ttgamma\ cross section is $\sigma(\ttgamma) = 2.4 \pm 0.2 \mbox{(stat.)} \pm 0.6 \mbox{(syst.)}$ pb for a photon with $E_{T} > 20$~GeV \cite{CMS:ttgamma}, with background modelling as the dominant systematic uncertainty. All measurements are in agreement with the NLO predictions.

\section{\ttv\ measurements}
The first step towards the measurement of the $t\bar{t}Z$ coupling at hadron colliders is the observation of the associated production of a $Z$ boson and top quark pair. The $t\bar{t}Z$ process can include the $Z$ boson as initial state radiation (ISR), i.e. radiated from the incoming quarks, or as final state radiation (FSR), i.e. radiated from the top quark. Only the latter processes are sensitive to the weak neutral current top coupling. In analogy to the $t\bar{t}\gamma$ processes, an off-shell $Z$ can also be radiated from the top quark decay products, although this contribution is expected to be negligible.
In contrast to $t\bar{t}Z$, the associated $W$ boson in $t\bar{t}W$ does not couple to the top quark, but is radiated from the incoming quarks (ISR process). The ISR processes are similar for $t\bar{t}Z$ and $t\bar{t}W$, and therefore, the understanding of the $t\bar{t}W$ production could be useful to disentangle the ISR from the FSR contribution in the $t\bar{t}Z$ processes. 

The first searches of the $t\bar{t}Z/W$ processes were performed by the ATLAS and CMS collaborations at $\sqrt{s}=7$ TeV using the full data set corresponding to an integrated luminosity of 4.7 \ifb\ and 5.0 \ifb, respectively. The ATLAS measurement searched for the \ttz\ process in final states with three leptons, setting an upper limit on the \ttz\ cross section of $\sigma_{\ttz}<0.71$ pb at 95\% CL \cite{ATLAS:ttZ7TeV}. The CMS collaboration measured the \ttz\ and \ttv\ (combined \ttz\ and \ttw) cross sections in final states with three leptons and with two leptons of same-sign electric charge, respectively, resulting in: $\sigma_{\ttz} = 0.28^{+0.14}_{-0.11}\mbox{(stat.)}^{+0.06}_{-0.03}\mbox{(syst.)}$ pb and $\sigma_{\ttv} = 0.43^{+0.17}_{-0.15}\mbox{(stat.)}^{+0.09}_{-0.07}\mbox{(syst.)}$ pb. These measurements have an observed significance of 3.3 and 3.0 standard deviations from the background-only hypothesis, respectively, providing first evidence of the \ttz\ process \cite{CMS:ttZ7TeV}. The measured cross sections are compatible within uncertainties with the corresponding NLO QCD predictions.

Both ATLAS and CMS collaborations have performed measurements of \ttz, \ttw\ and \ttv\ at $\sqrt{s}=8$ TeV, using the full data set corresponding to an integrated luminosity of 20.3 \ifb\ and 19.5 \ifb, respectively. More final state channels were included in the analysis than in the 7 TeV measurements, summarised in Table~\ref{tab:summarychannelsttv}. 

\begin{table}[h!]
\footnotesize
\centering     
\begin{tabular}{|c|c|c|c|c|c|c|}                                   
\hline
& \multicolumn{2}{c|}{\textbf{Trilepton}} &  \textbf{Same-sign (SS)} & \textbf{Four} & \multicolumn{2}{c|}{\textbf{Opposite-sign (OS)}} \\
& \multicolumn{2}{c|}{}  & \textbf{dilepton} & \textbf{lepton} & \multicolumn{2}{c|}{\textbf{dilepton}}  \\
\hline
\textbf{Channels}   & \threez & \threezveto & \twoss & \fourlep  & \rone & \rtwo \\
\hline
\textbf{Signal}     & \ttz & \ttw & \ttw & \ttz & \ttz\ and \ttw & \ttz \\
\hline
\textbf{Main Bkgds}  &  $tZ, WZ,$ & \ttz, \tth, & charge misID,  & $ZZ$, & \ttbar+jets & $Z$+jets \\ \cline{2-5}       
 &  \multicolumn{4}{c|}{lepton misID} & & \\
 \hline
\ms
\hline
\textbf{Fit regions}: & \multicolumn{6}{c|}{} \\
\hline
     & \threetwotri & \twothreetwotri  &  ($\mu\mu$) in  & & \threeonetwo & \threetwo \\
ATLAS & \fouronetri  &                 &  \twotwotri  & - & \fouronetwo  & \fourtwo \\
	  & \fourtwotri  &                 &  &  & \fiveonetwo  & \fivetwo \\
\hline
    & \fourtwotri  & \textit{[included}     & $(e^{\pm}e^{\pm}),(\mu^{\pm}\mu^{\pm}),$ & ($1 {\rm b}$) & & \\
CMS &              &  \textit{in \twoss]}   & $(e^{\pm}\mu^{\pm})$  in    & ($2 {\rm b}$) & - & - \\
	&              &              & \threeonetriCMS   & & & \\
 \hline
\end{tabular}
\caption{Overview of final state channels used in the ATLAS and CMS \ttv\ production cross section measurements at $\sqrt{s}=8$ TeV. The suffices ``Z" and ``Zveto" refer to the lepton invariant mass requirement $|\mll - m_Z| < 10$ GeV and $|\mll - m_Z| > 10$ GeV, respectively, where in the trilepton case, the \mll\ is the invariant mass of the same-flavour opposite-sign lepton pair. The fit regions are split into jet multiplicity, denoted by ``j'', and $b$-jet multiplicity, denoted by ``b''. Regions are exclusive unless given with a ``$\geq$''.} 
\label{tab:summarychannelsttv}                                                               
\end{table} 

 Both ATLAS and CMS perform a combination of the channels and measure \ttz, \ttw\ and \ttv\ production cross sections, by performing a profile likelihood fit. ATLAS measures cross sections relative to the NLO prediction: for \ttz: $\mu_{\ttz} = \sigma_{\ttz}/\sigma_{\ttz}^{SM} = 0.73^{+0.29}_{-0.26}$, for \ttw: $\mu_{\ttw} = 1.25^{+0.57}_{-0.48} $, and for \ttv: $\mu_{\ttv} = 0.89^{+0.23}_{-0.22}$. These measurements correspond to an observed 3.2, 3.1 and 4.9 standard deviations excess over the background-only hypothesis, respectively~\cite{ATLAS:ttZW8TeV}. CMS measures a \ttz\ cross section of $\sigma_{\ttz} = 200^{+80}_{-70}\mbox{(stat.)}^{+40}_{-30}\mbox{(syst.)}$ fb, a \ttw\ cross section of $\sigma_{\ttw} = 170^{+90}_{-80}\mbox{(stat.)} \pm 70\mbox{(syst.)}$ fb, and a \ttv\ cross section of $\sigma_{\ttv} = 380^{+100}_{-90}\mbox{(stat.)}^{+80}_{-70}\mbox{(syst.)}$~fb. These measurements have an observed significance of 3.1, 1.6 and 3.7 standard deviations from the background-only hypothesis, respectively~\cite{CMS:ttZW8TeV}. Besides the measurements of each process individually, a simultaneous measurement of \ttz\ and \ttw\ is also performed by both collaborations, and the results are shown in Figure~\ref{simultTTV8TeV}. These simultaneous measurements indicate a small correlation between \ttz\ and \ttw\ production cross sections. All cross section measurements are statistically limited and compatible with the NLO QCD predictions within uncertainties.
 
 \begin{figure}[h]
 \centering
 \includegraphics[width=18pc]{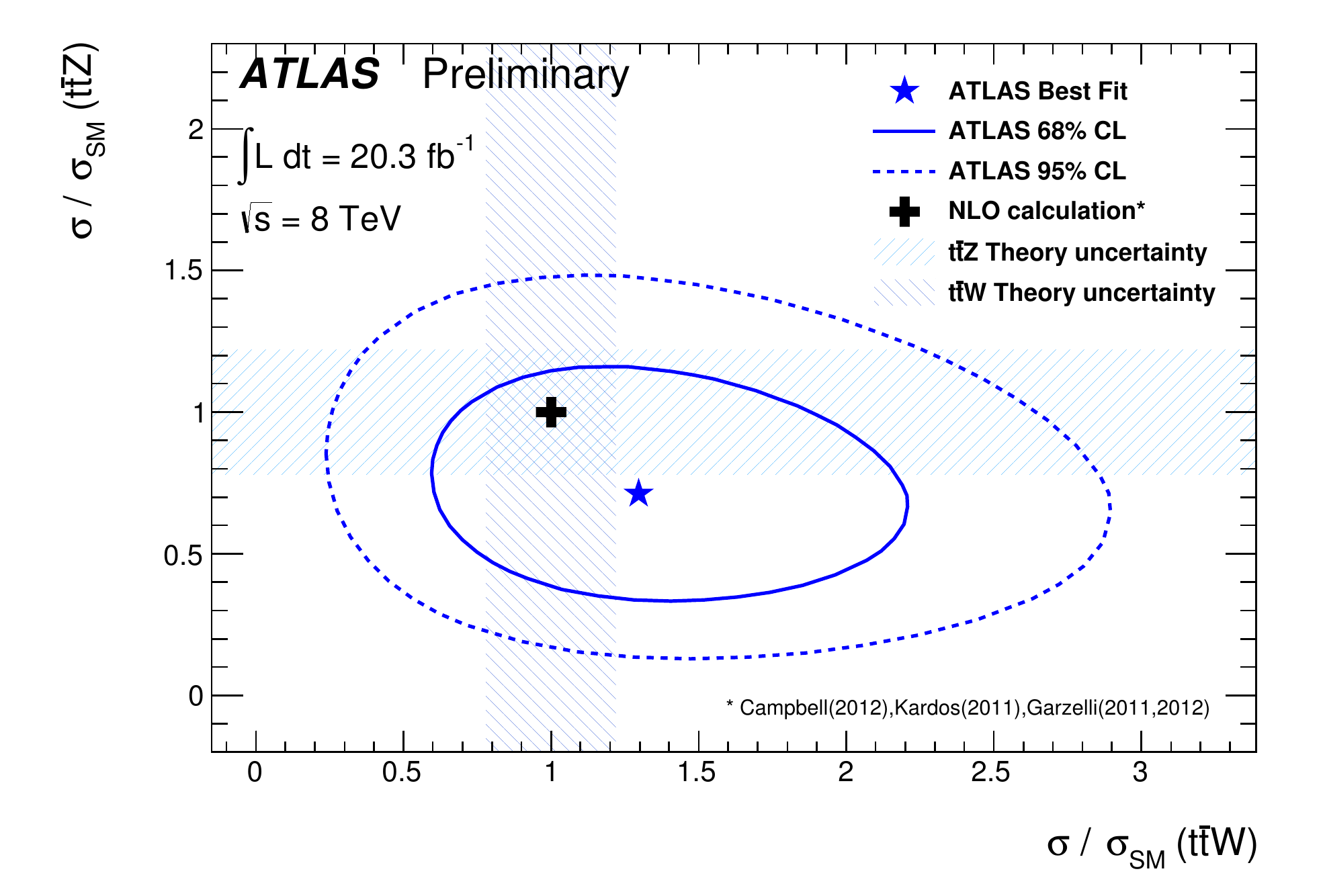}
 \includegraphics[width=17pc]{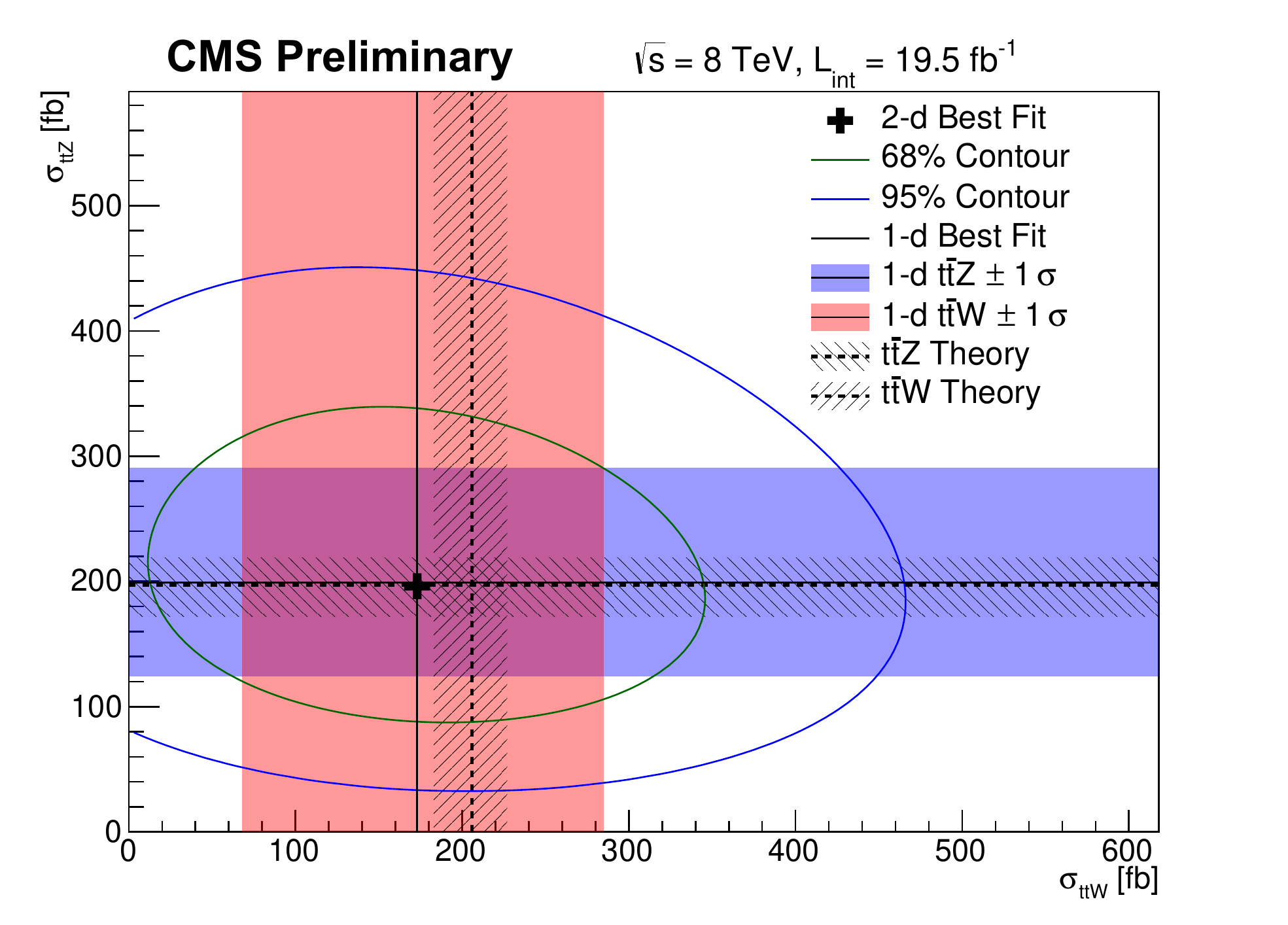}
 \caption{\label{simultTTV8TeV}Simultaneous measurement of \ttz\ and \ttw\ production cross section at $\sqrt{s}=8$ TeV by the ATLAS (left) and CMS (right) Collaborations \cite{ATLAS:ttZW8TeV,CMS:ttZW8TeV}.}
 \end{figure}

\section{Conclusions}

Almost 20 years after the discovery of the heaviest known quark, there are finally enough top quark pairs collected to explore the couplings of the top quark with the $\gamma,Z$ and $W$-boson via the production rate, or via the top quark decay for the study of the $Wtb$ vertex.
Evidence for both \ttz\ and \ttw\ production at $\sqrt{s}=8$~TeV has been recently announced, and the $Wtb$ coupling study improved with the latest precision \ttbar\ and single top measurements. No deviations from the SM predictions have been observed so far. However, this is just the beginning of such studies and more precise results are expected at higher luminosity and higher energy.

\end{document}